\documentclass[aps,pre,twocolumn,superscriptaddress,showpacs,floatfix]{revtex4-1}
\usepackage{hyperref}
\usepackage{amsmath}
\usepackage{graphicx}

\begin{document}

\title{Kernel polynomial representation of imaginary-time Green's functions}
\author{Li Huang}
\affiliation{ Beijing National Laboratory for Condensed Matter Physics, 
              and Institute of Physics, 
              Chinese Academy of Sciences, 
              Beijing 100190, 
              China }
\affiliation{ Science and Technology on Surface Physics and Chemistry Laboratory, 
              P.O. Box 718-35, 
              Mianyang 621907, 
              Sichuan, 
              China }

\author{Liang Du}
\affiliation{ Beijing National Laboratory for Condensed Matter Physics, 
              and Institute of Physics, 
              Chinese Academy of Sciences, 
              Beijing 100190, 
              China }

\date{\today}

\begin{abstract}
Inspired by the recent proposed Legendre orthogonal polynomial representation of imaginary-time Green's functions,
we develop an alternate representation for the Green's functions of quantum impurity models and combine it with 
the hybridization expansion continuous-time quantum Monte Carlo impurity solver. This representation is based on 
the kernel polynomial method, which introduces various integral kernels to filter fluctuations caused by the 
explicit truncations of polynomial expansion series and improve the computational precision significantly. As an 
illustration of the new representation, we reexamine the imaginary-time Green's functions of single-band Hubbard 
model in the framework of dynamical mean-field theory. The calculated results suggest that with carefully chosen 
integral kernels the Gibbs oscillations found in previous orthogonal polynomial representation have been suppressed 
vastly and remarkable corrections to the measured Green's functions have been obtained. 
\end{abstract}


\maketitle

\section{introduction}
\label{sec:intro}

The rapid development of efficient numerical and analytical methods for solving quantum impurity 
models has been driven in recent years by the great success of dynamical mean-field theory 
(DMFT)\cite{antoine:13,kotliar:865,held:2007} and its non-local extensions.\cite{maier:1027,toschi:045118,
rubtsov:033101} In the framework of DMFT, the momentum dependence of self-energy is neglected, 
then the solution of general lattice model may be obtained from the solution of an appropriately 
defined quantum impurity model plus a self-consistency condition. To solve the quantum impurity models
numerous quantum impurity solvers have been developed in the last decades.\cite{antoine:13,kotliar:865} 
In particular, continuous-time quantum Monte Carlo impurity solvers\cite{werner:076405,werner:155107,
haule:155113,gull:2008,rubtsov:035122} have become a very important tool for studying quantum impurity 
models, due to their accuracy, efficiency and ability to treat extreme low temperature and arbitrary 
interaction terms (for a recent review, see Ref.\onlinecite{gull:349}).

Among various continuous-time quantum Monte Carlo algorithms, the hybridization expansion version 
(abbreviated CT-HYB)\cite{werner:076405,werner:155107,haule:155113} is the most powerful and reliable 
impurity solver up to now and is widely used. In practice, as for the CT-HYB quantum impurity solver 
several severe technical limits still remain. One well-known problem is the high frequency noises 
commonly observed in the Matsubara Green's function and the self-energy.\cite{gull:235123,blumer:205120} 
Similar problems also arise in the calculations of imaginary-time Green's functions and vertex functions, 
which are less emphasized in the literatures. In order to cure these problems, intuitive idea is to run 
more statistics in the Monte Carlo simulations to suppress the data fluctuations as far as possible. 
This strategy should mitigate the problems, but it will not avert them and will deteriorate the efficiency 
of CT-HYB quantum impurity solver rapidly. Recently, Lewin Boehnke \emph{et al.}\cite{ortho:075145} 
have suggested to measure the single-particle and two-particle Green's function in the basis of 
Legendre orthogonal polynomials. The orthogonal polynomial method (OPM) provides a more compact 
representation of Green's functions than standard Matsubara frequencies, and therefore significantly 
reduces the memory storage size of these quantities. Moreover, it can be used as an efficient noise 
filter for various physical quantities within the CT-HYB quantum impurity solver: the statistical noise 
is mostly carried by high-order Legendre coefficients which should be truncated, while the physical 
properties are determined by low-order coefficients which should be retained. By and by the OPM is 
used for the computations of single-particle Green's function and lattice susceptibilities in the 
context of realistic DMFT calculations in combination with the local density approximation to the 
density functional theory (LDA+DMFT).\cite{deng:125137,boe:115128} By using the Legendre orthogonal 
polynomial representation, the accuracy of CT-HYB impurity solver is greatly improved. But according 
to careful examinations, the so-called Gibbs oscillations can be easily found in the resulting Green's 
functions and other physical quantities, which may be mainly due to the rough truncation of Legendre 
basis. The sign of Gibbs oscillations is that the resulting physical quantities are smooth but oscillating 
periodically with the scattered direct measurements. The situation is even worse in the insulating 
state where the Gibbs oscillations will cause the reconstructed single-particle Green's function to 
break the causality. Noted that a common procedure to damp these oscillations relies on an appropriate 
modification of the expansion coefficients by some integral kernels, which is the well-known kernel 
polynomial representation.\cite{wei:275} Thus we adopt the kernel polynomial method (KPM) to improve 
the measurement of single-particle and two-particle quantities within CT-HYB quantum impurity solver 
and expect to obtain significant improvements.

The rest of this paper is organized as follows: In Sec.\ref{subsec:ortho} a brief introduction 
to the orthogonal polynomial representation is provided. The original implementation is based
on Legendre polynomials only, and a straightforward generalization to Chebyshev polynomials is proposed.
In Sec.\ref{subsec:kernel} the kernel polynomial representation is presented in details. Then in 
Sec.\ref{sec:ben} we benchmark the KPM by reexamining the imaginary-time Green's function of 
single-band half-filled Hubbard model, and the characteristics of different integral kernel functions
which are used to alter the expansion coefficients are discussed. Section \ref{sec:con} serves 
as a conclusion and outlook. Finally in Appendix \ref{app:che}, concise introductions for the 
Chebyshev and Legendre orthogonal polynomial series are available as well. 

\section{method}
\label{sec:method}

\subsection{Orthogonal polynomial representation}
\label{subsec:ortho}

In the OPM, the imaginary-time Green's function $G(\tau)$ where $\tau \in [0,\beta]$ can 
be expanded in terms of Legendre orthogonal polynomials $P_{n}(x)$ defined on the interval $[-1,1]$,
\begin{equation}
\label{eq:gtau_leg}
G(\tau) = \frac{1}{\beta} \sum^{n_{\text{max}}}_{n \geq 0 } \sqrt{2n+1} P_{n}[x(\tau)] G_{n},
\end{equation}
\begin{equation}
\label{eq:g_leg_formal}
G_{n} = \sqrt{2n + 1} \int^{\beta}_{0} \text{d} \tau P_{n} [x(\tau)] G(\tau),
\end{equation}
where $\beta$ is inverse temperature, $x(\tau) = \frac{2\tau}{\beta} - 1$ and $G_{n}$ denotes 
the expansion coefficients of $G(\tau)$ in the Legendre orthogonal polynomials basis.\cite{ortho:075145} 
Since the expansion coefficients generally show a very fast decay with $n$, the expansion in 
Legendre polynomials can be truncated at a maximum order $n_{\text{max}}$. In the CT-HYB 
quantum impurity solver, the formula for measuring imaginary-time Green's function 
$G(\tau)$\cite{werner:076405,werner:155107,haule:155113} is
\begin{equation}
\label{eq:gtau_measure1}
G(\tau) = -\frac{1}{\beta}\left\langle \sum^{k}_{i=1} \sum^{k}_{j=1} 
\mathcal{M}_{ji} \Delta(\tau, \tau^{e}_{i} - \tau^{s}_{j})\right \rangle,
\end{equation}
\begin{equation}
\label{eq:gtau_measure2}
\Delta(\tau,\tau^{\prime}) =
\begin{cases}
\delta(\tau - \tau^{\prime}),          & \tau^{\prime} > 0\\
-\delta(\tau - \tau^{\prime} - \beta), & \tau^{\prime} < 0,
\end{cases}
\end{equation}
where $k$ is the order of diagrammatic perturbation expansion series, matrix element 
$(\mathcal{M}^{-1})_{ij} = F(\tau^{e}_{i} - \tau^{s}_{j})$ where $F(\tau)$ is the hybridization function,
$\tau^{e}_{i}$ and $\tau^{s}_{j}$ are the coordinates in imaginary-time axis for create and 
destroy operators, respectively. By utilizing Eq.(\ref{eq:gtau_measure1}) and Eq.(\ref{eq:gtau_measure2}), the 
Legendre coefficients for $G(\tau)$ finally become 
\begin{equation}
\label{eq:gl_measure}
G_{n} = -\frac{\sqrt{2n+1}}{\beta} \left\langle \sum^{k}_{i=1}\sum^{k}_{j=1} 
\mathcal{M}_{ji} \tilde{P}_{n}(\tau^{e}_{i} - \tau^{s}_{j} )\right\rangle,
\end{equation}
\begin{equation}
\tilde{P}_{n}(\tau) = 
\begin{cases}
P_{n}[x(\tau)],        & \tau > 0 \\
-P_{n}[x(\tau+\beta)], & \tau < 0.
\end{cases}
\end{equation}

Lewin Boehnke \emph{et al.}\cite{ortho:075145} have chosen the Legendre orthogonal polynomials 
as their preferred basis to expand single-particle and two-particle Green's functions. But it
should be stressed that \emph{a priori} different orthogonal polynomial bases may be used as 
well. Thus we try to generalize the OPM to use Chebyshev orthogonal polynomials as an optional 
basis. It is well-known that there exist two kinds of Chebyshev polynomials.\cite{is:2000} By 
using the Chebyshev polynomials of second kind $U_{n}(x)$ as basis, the imaginary-time Green's 
functions $G(\tau)$ can be expressed by the following equations,
\begin{equation}
\label{eq:gtau_che}
G(\tau) = \frac{2}{\beta} \sum^{n_{\text{max}}}_{n \geq 0} U_{n}[x(\tau)] G_{n},
\end{equation}
\begin{equation}
\label{eq:g_che_formal}
G_{n} = \frac{2}{\pi} \int^{\beta}_{0} \text{d} \tau U_{n} [x(\tau)] \sqrt{1 - x(\tau)^{2}}G(\tau).
\end{equation}
After a straightforward substitute, the Chebyshev coefficients for $G(\tau)$ finally become
\begin{equation}
\label{eq:gl_measure_che}
G_{n} = -\frac{2}{\pi\beta} 
\left\langle 
\sum^{k}_{i=1} \sum^{k}_{j=1} \mathcal{M}_{ji} 
\tilde{U}_{n}( \tau^{e}_{i} - \tau^{s}_{j} ) 
\sqrt{1 - \tilde{x}( \tau^{e}_{i} - \tau^{s}_{j} )^{2}}\right\rangle,
\end{equation}
where
\begin{equation}
\tilde{U}_{n}(\tau) = 
\begin{cases}
U_{n}[x(\tau)],        & \tau > 0 \\
-U_{n}[x(\tau+\beta)], & \tau < 0,
\end{cases}
\end{equation}
and
\begin{equation}
\tilde{x}(\tau) =
\begin{cases}
x(\tau),       & \tau > 0 \\
x(\tau+\beta), & \tau < 0.
\end{cases}
\end{equation}
The CT-HYB quantum impurity solver can directly accumulate the Legendre or Chebyshev coefficients 
$G_{n}$ instead of original Green's functions $G(\tau)$. Once the Monte Carlo sampling has been finished,
$G(\tau)$ can be reconstructed analytically by using the expansion coefficients. Since the coefficients 
decay very quickly, the orthogonal polynomial bases are much more compact and are particularly interesting 
for storing and manipulating the two-particle quantities, like vertex function etc. Furthermore, the Monte 
Carlo noises are mainly concentrated in the high-order expansion coefficients, and the numerical values of them 
are usually very small. So a rough truncation method can be developed to filter out the noises and obtain 
more smooth and accurate results.

\subsection{Kernel polynomial representation}
\label{subsec:kernel}

\begin{table*}
\centering
\caption{Summary of different integral kernel functions $f_{n}$ that can be used to improve the quality
of an order $N$ Chebyshev or Legendre series}
\label{tab:damping_kernel}
\begin{tabular}{llll}
\hline 
\hline
name & $f_{n}$ & parameters & positive \\
\hline
Jackson     & $\frac{1}{N} \left[(N-n+1)\cos(\frac{\pi n}{N+1}) + \sin(\frac{\pi n}{N+1})\cot(\frac{\pi}{N+1})\right]$ & none  & yes \\
Lorentz     & $\sinh[\lambda(1-n/N)]/\sinh(\lambda) $ &  $\lambda \in \mathcal{R}$ & yes \\
Fej\'{e}r       & $1 - n/N$ & none  & yes \\
Wang-Zunger & $\exp\left[-\left(\alpha\frac{n}{N}\right)^{\beta}\right]$ & $\alpha$, $\beta \in \mathcal{R}$ & no \\
Dirichlet   & 1 & none  & no \\
\hline
\hline
\end{tabular}
\end{table*}

\begin{figure}
\centering
\includegraphics[scale=0.60]{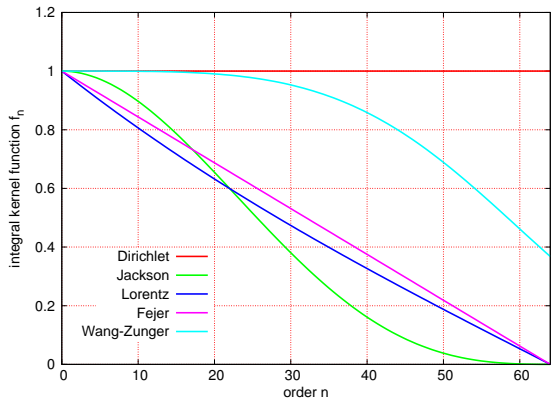}
\caption{(Color online) Classic integral kernels $f_{n}$ used to improve the quality of polynomial expansion 
series. In this figure, the order of expansion series is $N = 64$. The Dirichlet, Jackson, and Fej{\'e}r 
kernels take no parameters. For Lorentz kernel, the $\lambda$ parameter is fixed to be 1.0. And for Wang-Zunger 
kernel, the $\alpha$ and $\beta$ parameters are 1.0 and 4.0, respectively.}
\label{fig:ker}
\end{figure}

The basic idea of OPM is to expand single-particle Green's functions $G(\tau)$ in infinite series 
of Chebyshev or Legendre polynomials, and then use Monte Carlo algorithm to sample the expansion 
coefficients $G_{n}$ directly. As expected for a numerical approach, however, the expansion series 
will remain finite actually, and we thus arrive at a classical problem of approximation theory. In 
our case the problem is equivalent to find the best approximation to $G(\tau)$ given a finite number 
of $G_{n}$. Experience shows that a simple truncation of the infinite series leads to poor precision 
and fluctuations, which also known as Gibbs oscillations.\cite{wei:275} For examples, as for the 
reconstructed Green's function $G(\tau)$ in insulating state, almost periodic Gibbs oscillations are 
clearly identified in a wide $\tau$ range.

A common procedure to damp the Gibbs oscillations is to introduce some kind of integral kernel function $f_{n}$
and change the expansion coefficients from $G_{n}$ to $G_{n}f_{n}$.\cite{wei:275} Obviously, the simplest
integral kernel function, which is usually attributed to Dirichlet, is obtained by setting $f_{n} = 1$. By using
the Dirichlet kernel, the KPM is equivalent to previous OPM. In addition to the Dirichlet kernel, other 
classic integral kernel functions, like Jackson, Lorentz, Fej\'{e}r, and Wang-Zunger etc., are collected 
in Tab.\ref{tab:damping_kernel} and plotted in Fig.\ref{fig:ker} respectively. Note that for all the kernels
$f_{0}$ must be equal to 1 and $f_{1}$ must approach 1 as $n \rightarrow \infty$. The optimal integral kernel 
function partially depends on the considered application. According to the literature,\cite{wei:275} the 
Jackson kernel may be the best for most applications, the Lorentz kernel may be the best for Green's function, 
while the Fej\'{e}r kernel is mainly of academic interest. 

Finally, we note that the integral kernel functions $f_{n}$ can be evaluated and stored in advance, so the 
KPM has not effect on the computational efficiency of CT-HYB quantum impurity solver. The implementation of 
KPM is very simple, only small modifications are needed for the OPM's version of CT-HYB, i.e., replacing 
$G_{n}$ by $G_{n}f_{n}$.
Since the kernel polynomial and orthogonal polynomial representations are only alternate bases for single-particle 
and two-particle quantities, so both methods can be implemented in segment picture\cite{werner:076405} and 
general matrix\cite{werner:155107} formulations of CT-HYB impurity solvers to improve the accuracy and efficiency.

\section{benchmark}
\label{sec:ben}

In this section, we try to benchmark the kernel polynomial representation and compare the calculated 
results with those obtained by orthogonal polynomial representation and conventionally direct measurements. 
For the sake of simplicity, a single-band Hubbard model on Bethe lattice is used as a toy model to examine 
our implementations of OPM and KPM. Here $U = 4.0$ and $\beta = 10.0$ for metallic case, and $U = 6.0$ and 
$\beta = 50.0$ for insulating case. The band is with bandwidth 2.0, and a semicircular density of states is 
chosen. The chemical potential $\mu$ is fixed to be $U/2$ to keep the model under half-filling. Unless it 
is specifically stated, this model is used throughout this section. This toy model is studied in the framework 
of single site DMFT\cite{antoine:13,kotliar:865} and the segment picture version of CT-HYB\cite{werner:076405} 
is used as quantum impurity solver. In each DMFT iterations, typically $4 \times 10^{8}$ Monte Carlo sweeps 
have been performed to reach sufficient numerical precision.
 
\subsection{Metallic state}
\label{subsec:metal}

\begin{figure}
\centering
\includegraphics[scale=0.60]{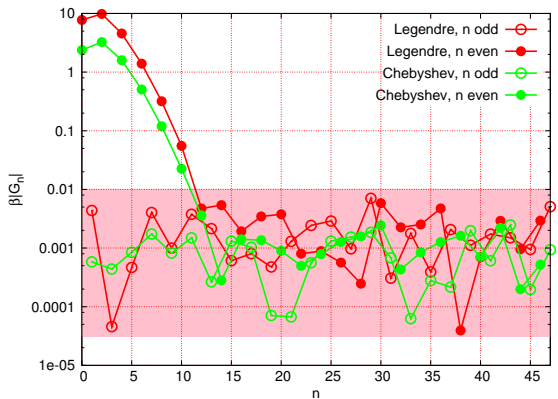}
\caption{(Color online) Chebyshev and Legendre coefficients $\beta |G_n|$ of 
imaginary-time Green's functions of the single-band half-filled Hubbard model on 
the Bethe lattice within DMFT. The Coulomb interaction strength $U$ is 4.0 and inversion
temperature $\beta$ is 10. The coefficients in the pink region contribute very 
little to the resulting Green's function. \label{fig:mc}}
\end{figure}

\begin{figure}
\centering
\includegraphics[scale=0.60]{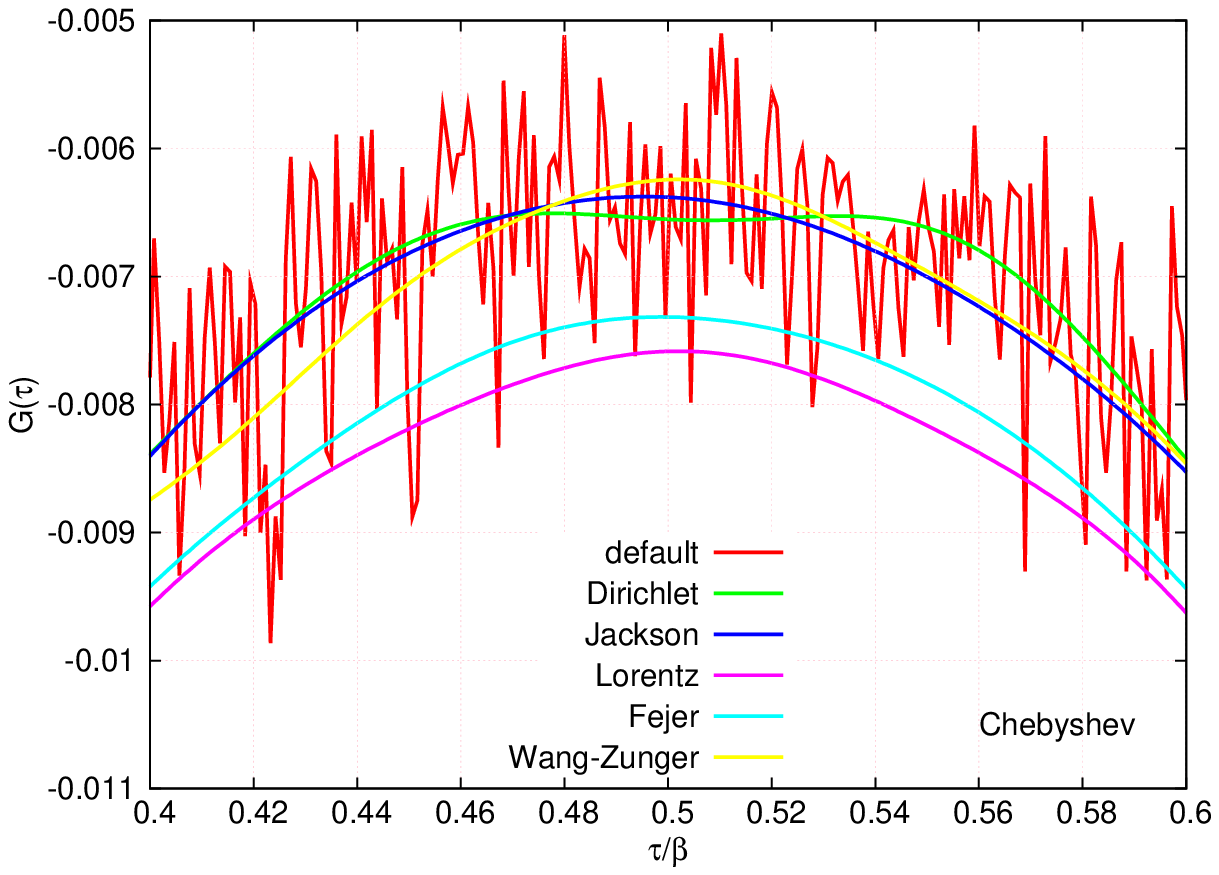}
\includegraphics[scale=0.60]{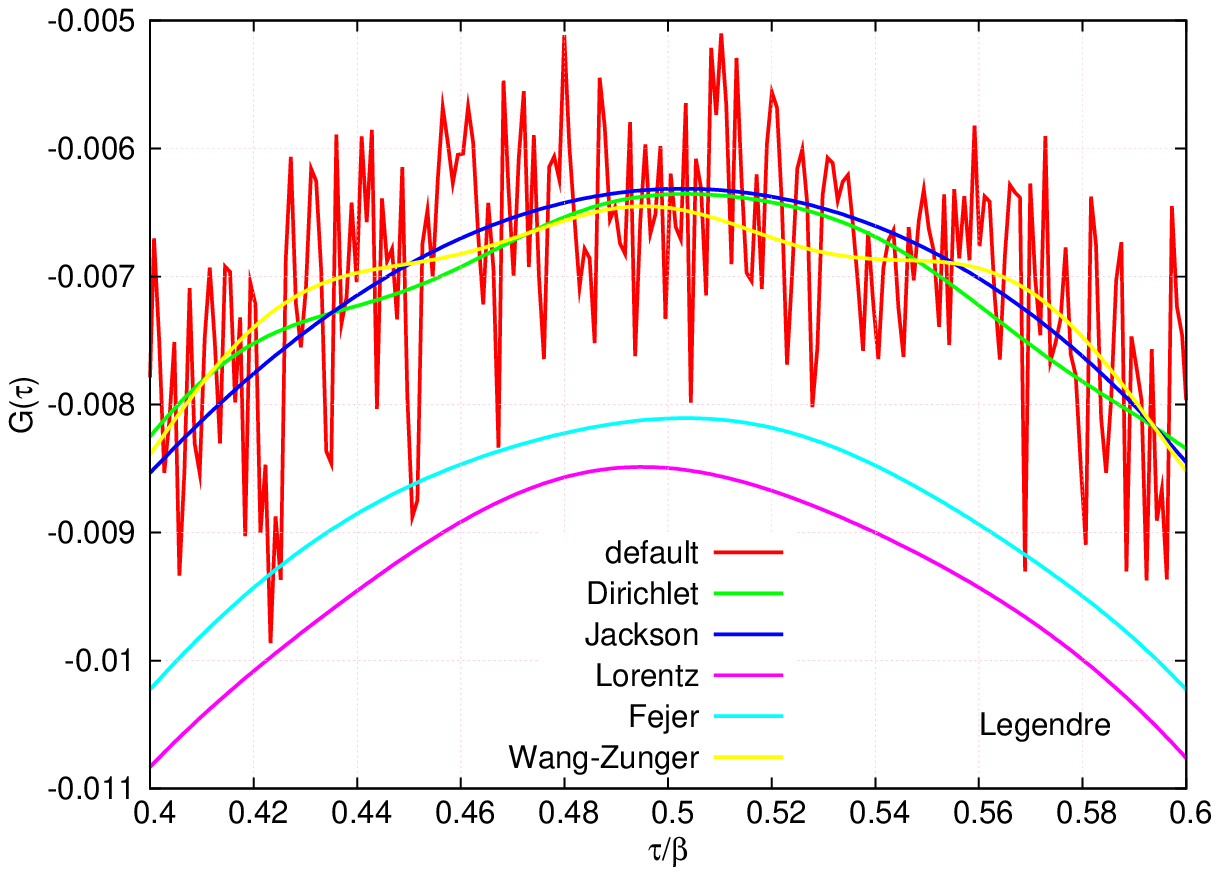}
\caption{(Color online) The imaginary-time Green's function $G(\tau)$ in $\tau \in 
[0.6\beta,0.8\beta]$ interval of the single-band half-filled Hubbard model. The 
Coulomb interaction strength $U$ is 4.0 and $\beta = 10$. Upper panel: $G(\tau)$ calculated 
by Chebyshev polynomials with or without integral kernel functions. Lower panel: $G(\tau)$ 
calculated by Legendre polynomials with or without integral kernel functions. The order for 
polynomial expansion series is 24.\label{fig:mg}}
\end{figure}

Let's first concentrate our attentions to the metallic state. In figure \ref{fig:mc} the ``bare" 
expansion coefficients $\beta|G_{n}|$ of Chebyshev and Legendre orthogonal polynomials are shown. 
As is pointed out by Lewin Boehnke \emph{et al.}\cite{ortho:075145}, due to the constraint of 
particle-hole symmetry the expansion coefficients for odd order $n$ should be zero. Indeed, the 
coefficients in our data for odd $n$'s all take on a very small value, compatible with a vanishing
value within their error bars. The even $n$ coefficients instead show a very fast decay. As is 
shown in this figure, $n_{\text{max}} = 15 \sim 25$ is enough for both Chebyshev and Legendre 
polynomial representations to obtain converged and accurate results. In our simulations, 
$n_{\text{max}}$ is fixed to be 24.

Figure \ref{fig:mg} shows the calculated imaginary-time Green's function $G(\tau)$ by using KPM with 
different orthogonal polynomials and integral kernel functions. It is apparent that the directly 
measured $G(\tau)$ is full of noises and fluctuations, which are negative for the later analytical 
continuation procedure.\cite{jarrell:133} Once the OPM is used (i.e., Dirichlet kernel $f_{n} = 1$ 
is adopted), $G(\tau)$ turns smooth but obvious undulations still exist. If the Lorentz, Fej\'{e}r, 
and Wang-Zunger kernels are applied one by one, the Green's functions are smooth and without obvious
undulations, but deviate systematically from the scattered data. As a general view, the Jackson 
kernel function is the optimal choice. The resulting $G(\tau)$ evaluated by Jackson kernel function 
is smooth and nicely interpolates the directly measured data. As is expected, the type of orthogonal 
polynomials has little impact to the interpolated $G(\tau)$. It seems that the Chebyshev polynomials 
do a bit better than Legendre polynomials.

\subsection{Insulating state}
\label{subsec:insulator}

\begin{figure}
\centering
\includegraphics[scale=0.60]{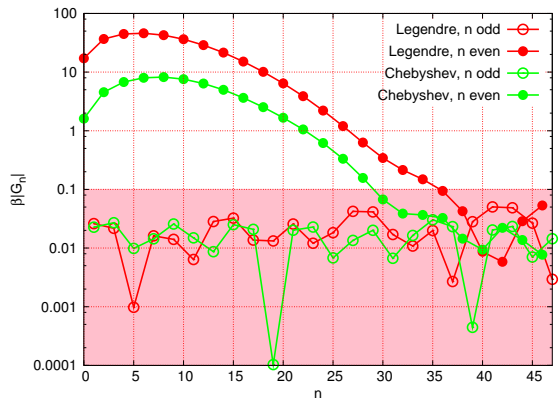}
\caption{(Color online) Chebyshev and Legendre coefficients $\beta |G_n|$ of 
imaginary-time Green's functions of the single-band half-filled Hubbard model on 
the Bethe lattice within DMFT. The Coulomb interaction strength $U$ is 6.0 and inversion
temperature $\beta$ is 50. The coefficients in the pink region contribute very 
little to the resulting Green's function. \label{fig:ic}}
\end{figure}

\begin{figure}
\centering
\includegraphics[scale=0.60]{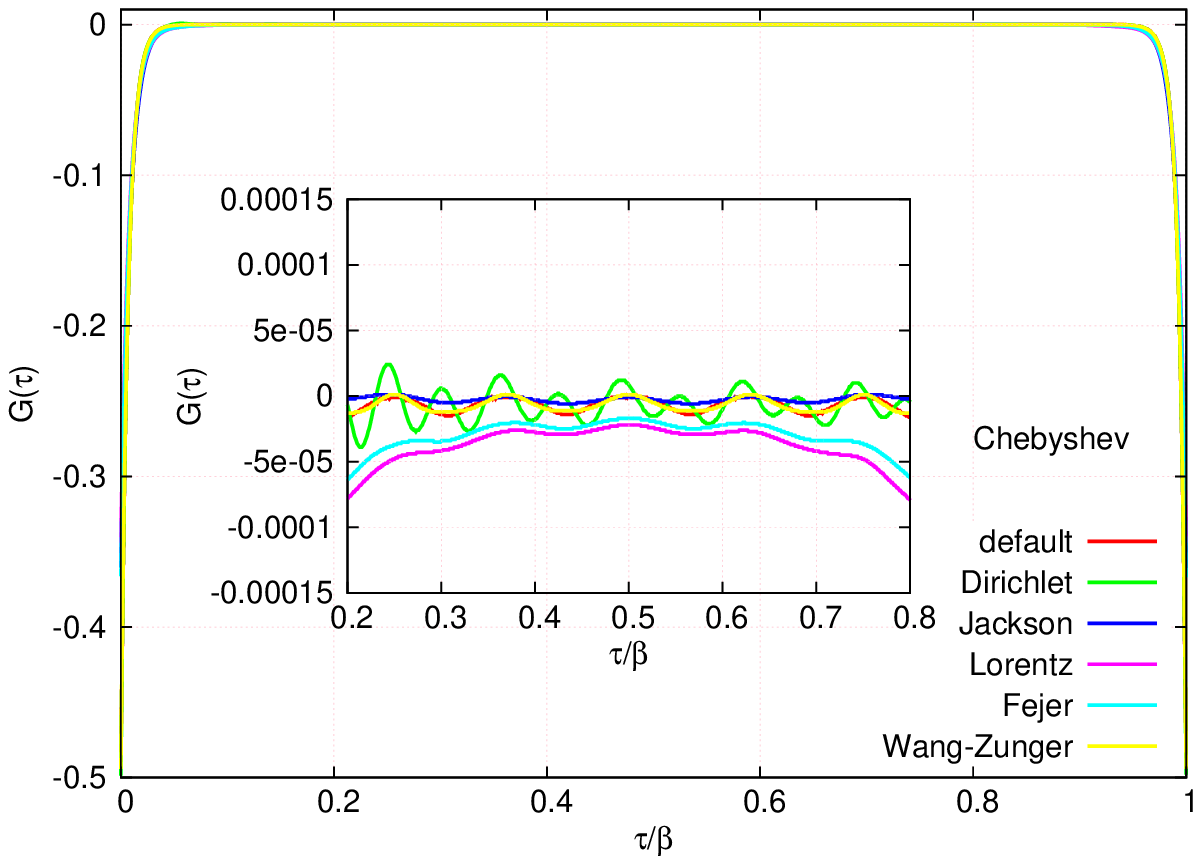}
\includegraphics[scale=0.60]{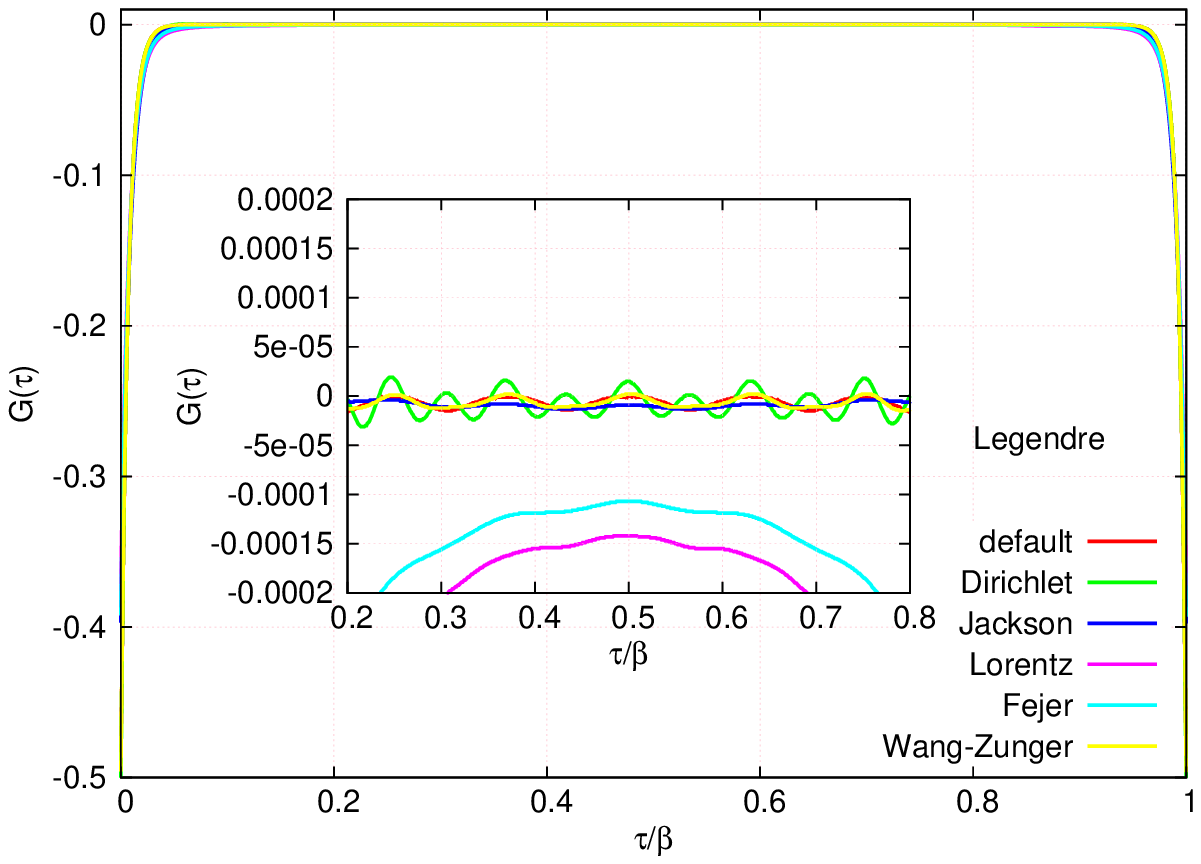}
\caption{(Color online) The imaginary-time Green's function $G(\tau)$ of the single-band 
half-filled Hubbard model. The Coulomb interaction strength $U$ is 6.0 and $\beta = 50$. 
Upper panel: $G(\tau)$ calculated by Chebyshev polynomials with or without integral kernel
functions. Lower panel: $G(\tau)$ calculated by Legendre polynomials with or without 
integral kernel functions. The insets in both panels show the fine structures of $G(\tau)$
in $\tau \in [0.2\beta,0.8\beta]$ interval. The order for polynomial expansion series is 
64.\label{fig:ig}}
\end{figure}

Now let's turn to the insulating state. When $U = 6.0$ and $\beta = 50$ an definitely
insulating solution is obtained within DMFT. The ``bare" Chebyshev and Legendre coefficients 
$\beta|G_{n}|$ of $G(\tau)$ are shown in Fig.\ref{fig:ic}. Just similar to the metallic state, 
$G_{n}$ takes very small value for odd $n$ and can be ignored safely, and for even $n$, $G_{n}$ converges
to zero very quickly. For Chebyshev and Legendre polynomials, $n_{\text{max}} = 35 \sim 45$ or 
$n_{\text{max}} = 40 \sim 50$, respectively. Thus the Chebyshev polynomials is more compact 
and efficient than Legendre polynomials. In current simulations, $n_{\text{max}}$ is fixed to 
be 64 uniformly.

The calculated imaginary-time Green's function $G(\tau)$ by using KPM with different orthogonal 
polynomials and integral kernel functions are illustrated in Fig.\ref{fig:ig}. A cursory look 
could lead you to believe that the reconstructed Green's functions by KPM or OPM agree rather 
well with the scattered data. Next let's zoom in $\tau \in [0.2\beta,0.8\beta]$ interval and 
check carefully the magnified $G(\tau)$, which are just depicted in the insets of Fig.\ref{fig:ig}. 
Clearly, $G(\tau)$ is very close to zero in this region. The scattered data obtained by direct 
measurement exhibit periodical undulations. The reconstructed $G(\tau)$ by OPM (with Dirichlet 
kernel) displays stronger periodical oscillations and violates the causality at the same time, 
which means that the results will be even deteriorated by using orthogonal polynomial representation. 
The results obtained by Wang-Zunger kernel fit original data very well and obvious improvement 
is not observed. As for the Lorentz and Fej\'{e}r kernels, the calculated results deviate the 
scattered data systematically. Again, the Jackson kernel is the optimal choice. The calculated 
results are very smooth, perfectly interpolate the scattered data, and obey the causality.

\section{conclusions}
\label{sec:con}

It is suggested that the OPM based on Legendre orthogonal polynomials can be used to improve 
the accuracy and computational efficiency of CT-HYB quantum impurity solver.\cite{ortho:075145} 
In this paper, we develop a better representation to calculate the single-particle and two-particle
quantities. Firstly, we generalize the OPM to Chebyshev orthogonal polynomial basis. Secondly, 
the KPM based on various integral kernel functions is proposed to damp the Gibbs oscillations 
observed in the single-particle and two-particle Green's functions obtained with OPM and improve the 
accuracy of them further. According to the benchmark results for single-band half-filled Hubbard model,
it is demonstrated that the Jackson kernel is the optimal choice for imaginary-time Green's function
$G(\tau)$ and other quantities. Though the KPM presented in this paper is mainly developed for 
the CT-HYB quantum impurity solver, it can be easily generalized to other continuous-time quantum 
Monte Carlo impurity solvers.\cite{gull:349}

\begin{acknowledgments}
We acknowledge financial support from the National Science Foundation ͑of China and that
from the 973 program of China under Contract No.2007CB925000 and No.2011CBA00108.
\end{acknowledgments}

\appendix
\section{Chebyshev and Legendre polynomials}
\label{app:che}

\begin{figure}
\centering
\includegraphics[scale=0.60]{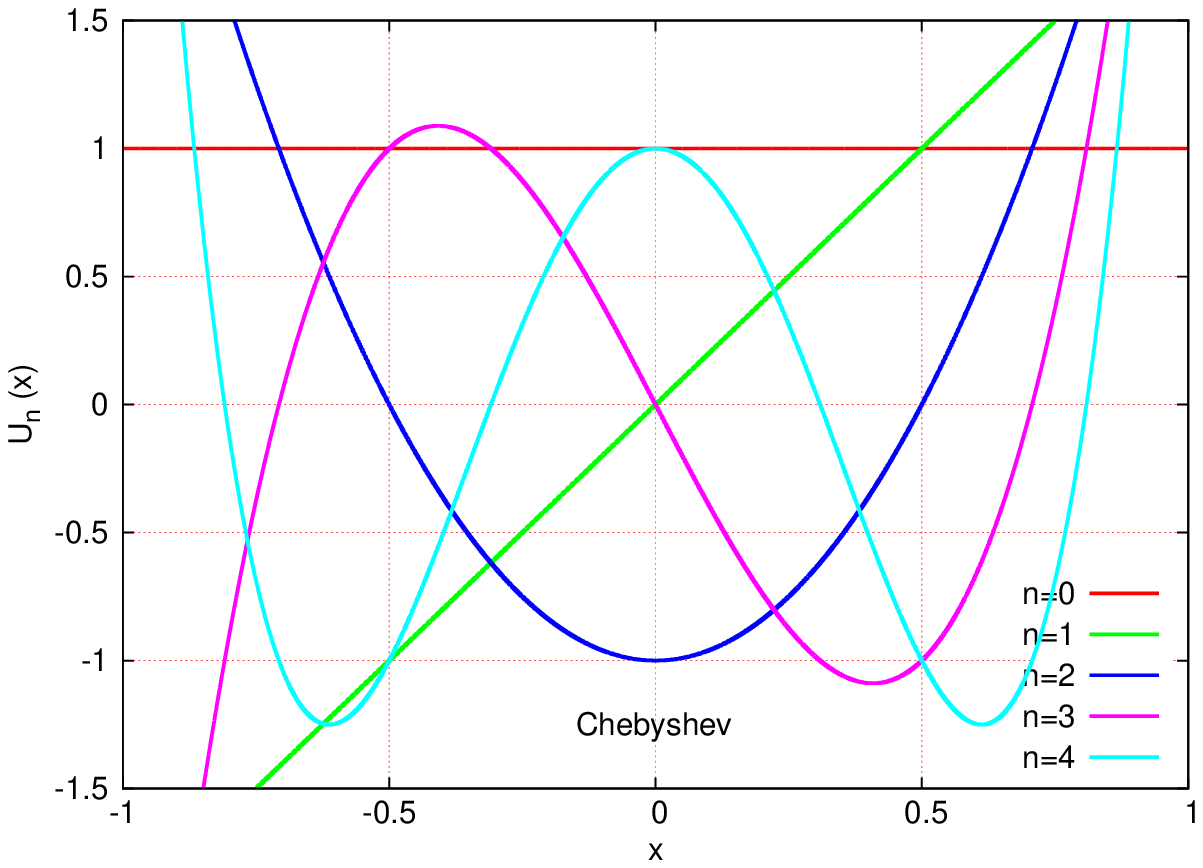}
\includegraphics[scale=0.60]{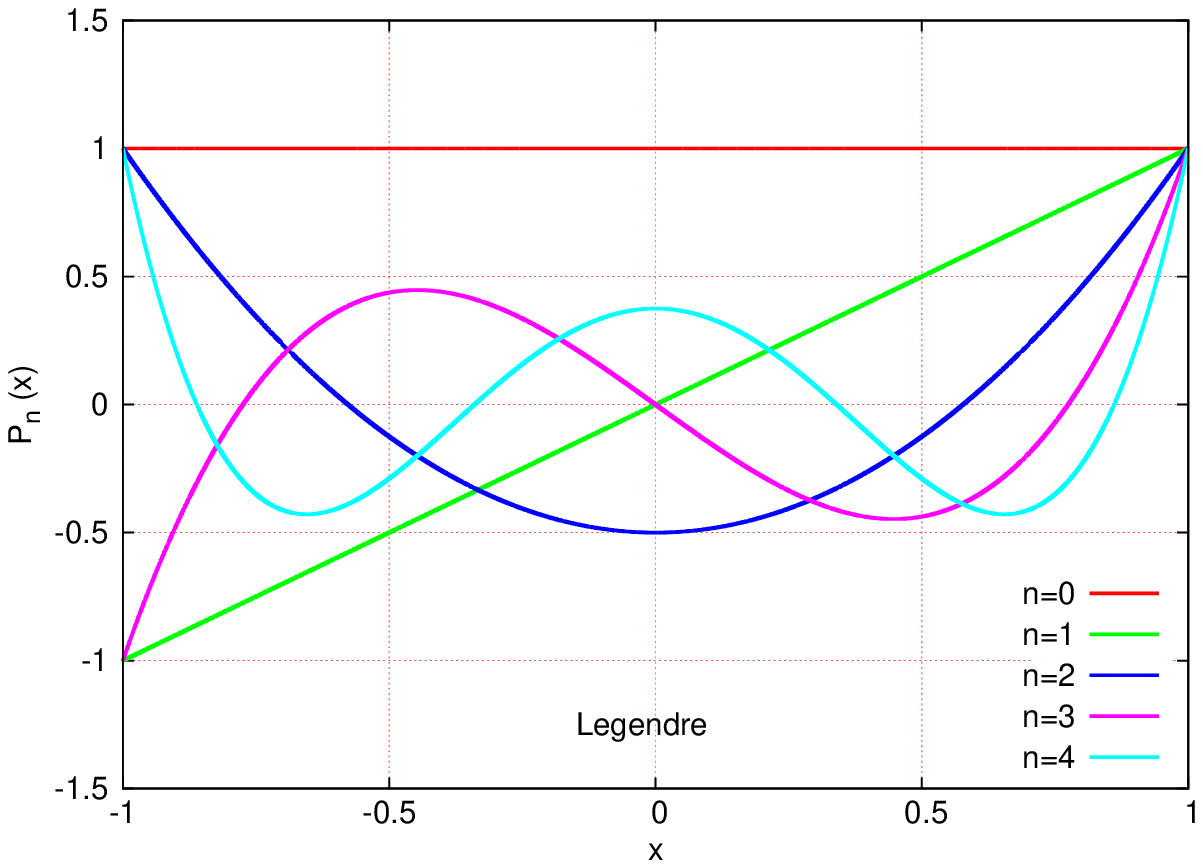}
\caption{(Color online) Chebyshev and Legendre orthogonal polynomials. Upper panel: The first 
few Chebyshev polynomials of the second kind $U_n$ in the domain $−1 < x < 1$. Lower panel: The 
first few Legendre polynomials $P_n$ in the domain $−1 < x < 1$. \label{fig:che_leg}}
\end{figure}

In this appendix, we summarize for convenience some basic properties of the Chebyshev and 
Legendre polynomials, and the first few Chebyshev and Legendre polynomials are illustrated 
in Fig.\ref{fig:che_leg}. Further reference can be found in Ref.\onlinecite{is:2000}.

The Chebyshev polynomials are a sequence of orthogonal polynomials which can be defined 
recursively. One usually distinguishes between Chebyshev polynomials of the first kind 
which are denoted by $T_{n}$ and Chebyshev polynomials of the second kind which are denoted 
by $U_{n}$. The Chebyshev polynomials $T_{n}$ or $U_{n}$ are polynomials of degree $n$. 
$T_{n}$ and $U_{n}$ are defined by the following recurrence relations:
\begin{equation} 
T_0(x) = 1,
\end{equation} 
\begin{equation}
T_1(x) = x,
\end{equation}
\begin{equation} 
T_{n+1}(x) = 2xT_n(x) - T_{n-1}(x). 
\end{equation} 
and
\begin{equation} 
U_0(x) = 1,
\end{equation}
\begin{equation} 
U_1(x) = 2x,
\end{equation}
\begin{equation} 
U_{n+1}(x) = 2xU_n(x) - U_{n-1}(x). 
\end{equation}
The Chebyshev polynomials of the first and second kind are closely related by the 
following equations:
\begin{equation}
T_{n+1}(x) = xT_n(x) - (1 - x^2)U_{n-1}(x), 
\end{equation}
and
\begin{equation}
T_n(x) = U_n(x) - x \, U_{n-1}(x).
\end{equation}
The Chebyshev polynomials of the first kind are orthogonal with respect to the weight 
$\frac{1}{\sqrt{1-x^{2}}}$ on the interval $[-1,1]$, i.e. we have:
\begin{equation}
\int_{-1}^1 T_n(x)T_m(x)\,\frac{dx}{\sqrt{1-x^2}}= 
\begin{cases} 0 &: n\ne m, \\ \pi &: n=m=0,\\ \pi/2 &: n=m\ne 0. \end{cases} 
\end{equation}
Similarly, the Chebyshev polynomials of the second kind are orthogonal with respect to 
the weight $\sqrt{1-x^2}$ on the interval $[−1,1]$, i.e. we have:
\begin{equation}
\int_{-1}^1 U_n(x)U_m(x)\sqrt{1-x^2}\,dx = 
\begin{cases} 0 &: n\ne m, \\ \pi/2 &: n=m. \end{cases} 
\end{equation}

Similar to the Chebyshev polynomials, Legendre polynomials $P_{n}$ are orthogonal polynomials 
of degree $n$, which can be defined by the following recurrence relations:
\begin{equation}
P_0(x) = 1,
\end{equation}
\begin{equation}
P_1(x) = x, 
\end{equation}
\begin{equation}
(n+1) P_{n+1}(x) = (2n+1) x P_n(x) - n P_{n-1}(x).
\end{equation}
The orthogonality of Legendre polynomials on the interval [-1,1] reads:
\begin{equation}
\int_{-1}^{1} P_m(x) P_n(x)\,dx = \frac{2} {2n + 1} \delta_{mn}. 
\end{equation}

\bibliography{kpm}

\end{document}